\documentclass[sigconf, authorversion]{acmart}

\usepackage{balance}
\usepackage{subcaption}
\usepackage{multirow}

\usepackage[version-1-compatibility]{siunitx}

\AtBeginDocument{%
  \providecommand\BibTeX{{%
    \normalfont B\kern-0.5em{\scshape i\kern-0.25em b}\kern-0.8em\TeX}}}

\copyrightyear{2022} 
\acmYear{2022} 
\setcopyright{acmlicensed}\acmConference[AutomotiveUI '22]{14th International Conference on Automotive User Interfaces and Interactive Vehicular Applications}{September 17--20, 2022}{Seoul, Republic of Korea}
\acmBooktitle{14th International Conference on Automotive User Interfaces and Interactive Vehicular Applications (AutomotiveUI '22), September 17--20, 2022, Seoul, Republic of Korea}
\acmPrice{15.00}
\acmDOI{10.1145/3543174.3546840}
\acmISBN{978-1-4503-9415-4/22/09}

\begin{document}


\title[What's on your mind? A Mental and Perceptual Load Estimation Framework]{What's on your mind? A Mental and Perceptual Load Estimation Framework towards Adaptive In-vehicle Interaction while Driving}


\author{Amr Gomaa}
\affiliation{%
  \institution{German Research Center for Artificial Intelligence (DFKI)}
  \city{Saarbr{\"u}cken}
  \country{Germany}
}
\email{amr.gomaa@dfki.de}

\author{Alexandra Alles}
\affiliation{%
  \institution{German Research Center for Artificial Intelligence (DFKI)}
  \city{Saarbr{\"u}cken}
  \country{Germany}
}
\email{alexandra_katrin.alles@dfki.de}

\author{Elena Meiser}
\affiliation{%
  \institution{German Research Center for Artificial Intelligence (DFKI)}
  \city{Saarbr{\"u}cken}
  \country{Germany}
}
\email{elena.meiser@dfki.de}

\author{Lydia Helene Rupp}
\affiliation{%
  \institution{German Research Center for Artificial Intelligence (DFKI)}
  \city{Saarbr{\"u}cken}
  \country{Germany}
}
\email{lydia_helene.rupp@dfki.de}

\author{Marco Molz}
\affiliation{%
  \institution{German Research Center for Artificial Intelligence (DFKI)}
  \city{Saarbr{\"u}cken}
  \country{Germany}
}
\email{marco.molz@dfki.de}

\author{Guillermo Reyes}
\affiliation{%
  \institution{German Research Center for Artificial Intelligence (DFKI)}
  \city{Saarbr{\"u}cken}
  \country{Germany}
}
\email{guillermo.reyes@dfki.de}

\renewcommand{\shortauthors}{Gomaa et al.}

\begin{abstract}

Several researchers have focused on studying driver cognitive behavior and mental load for in-vehicle interaction while driving. Adaptive interfaces that vary with mental and perceptual load levels could help in reducing accidents and enhancing the driver experience.
In this paper, we analyze the effects of mental workload and perceptual load on psychophysiological dimensions and provide a machine learning-based framework for mental and perceptual load estimation in a dual task scenario for in-vehicle interaction~(\url{https://github.com/amrgomaaelhady/MWL-PL-estimator}). We use off-the-shelf non-intrusive sensors that can be easily integrated into the vehicle's system. Our statistical analysis shows that while mental workload influences some psychophysiological dimensions, perceptual load shows little effect. Furthermore, we classify the mental and perceptual load levels through the fusion of these measurements, moving towards a real-time adaptive in-vehicle interface that is personalized to user behavior and driving conditions. We report up to 89\% mental workload classification accuracy and provide a real-time minimally-intrusive solution.

\end{abstract}




\maketitle
\pagestyle{plain}

\section{Introduction}

Driving is an everyday task in which lapses in attention can have fatal consequences. The driving task is still one of the most complex tasks for the human brain despite large advancements in the automotive industry~\cite{groeger2000understanding,Gabaude2012}; drivers perform several visual and auditory mental sub-tasks to be able to drive adequately, such as perception, expectation, judgment, planning, and execution~\cite{FASTENMEIER2007drivingtaskanalysis}. According to the WHO, approximately $1.3$ million people die each year as a result of road traffic accidents\footnote{\url{https://www.who.int/news-room/fact-sheets/detail/road-traffic-injuries}}. Since almost half of all vehicular accidents involve driver inattention~\cite{stutts2001role}, driving safety could benefit from advances in attention research.
Additionally, the number of available secondary tasks and therefore the number of distractions rises with the advancement of in-vehicle human-machine interfaces (HMI)~\cite{zwahlen1988safety,Vicente2015DriverSystem}. 
Thus, recent research advancements focus on designing interfaces with minimal interference on the driving task, while providing features aiding the driver's comfort and safety~\cite{Olaverri-Monreal2014CapturingExpectations,Gibson2016User-centeredConsoles,Carsten2019HowSolutions}.
One way of evaluating in-vehicle interfaces is measuring the mental effort they put on the driver, specifically, perceptual load (PL) and mental workload (MWL). MWL has been a topic of interest for the research community for the last 50 years, since Casali et al.'s work on defining and estimating mental effort~\cite{casali1983comparison}. PL has been discussed since the mid 1980s and solved a long-standing debate between early and late selection in visual attention processes~\cite{kahneman1984changing}. Both MWL and PL play an important role in driver safety and the driving experience in terms of auditory and visual distractions.
In this paper, we analyze different methods for objectively estimating the driver's MWL and PL (as separate constructs) while performing the dual task of controlling a secondary system while driving. We provide an end-to-end machine learning (ML) based framework for MWL and PL estimation separately using non-intrusive easily integrable sensors that are in concurrence with current sensors in modern vehicles~\cite{2019BMW2019,2020Toyota2020,2021Mercedes-Benz2021}.



\section{Background and Related Work}

\subsection {Mental Workload (MWL)}
Mental workload is an important factor in human-machine interaction and has often been studied in the context of driving~\cite{boer2001behavioral,Caird2018DoesStudies,horrey2007vehicle,Lohani2019ADriving,Marquart2015ReviewWorkload,Cantin2009MentalComplexity,Verwey2000On-lineMeasures}. Although no universal definition of MWL has been proposed to date, several authors agree that MWL is an interaction of task demands, other environmental factors, and human characteristics (e.g., attentional capacities or task experience)~\cite{Marquart2015ReviewWorkload, tao2019systematic}. Hart and Staveland~\cite{hart1988development} state that ``workload is a hypothetical construct that represents the cost incurred by a human operator to achieve a particular level of performance''.
This means that a driver is expending mental effort to maintain a subjectively safe driving behavior~\cite{boer2001behavioral}, while various mediating factors are continuously changing (e.g., different road conditions~\cite{di2018eeg}). There have been several attempts to explain MWL, however, the multiple resource theory by Wickens~\cite{wickens2008multiple} is the most widely used approach. It assumes that several modality-specific attentional resources are used when performing a mentally demanding task. In dual task scenarios, several mental resources (e.g., visual or auditory) can process information simultaneously. This suggests that driving could be defined mainly as a mental, visual, and physical task~\cite{Marquart2015ReviewWorkload}. One task commonly performed secondary to driving is talking and dialing on a phone or talking to a passenger. The latter task uses more auditory and verbal resources, while the former requires visual and manual resources~\cite{Caird2018DoesStudies}. Caird et al.~\cite{Caird2018DoesStudies} showed that dialing a number has highly detrimental effects on driving performance increasing accidents risk~\cite{horrey2007vehicle}. A performance loss may even occur when the secondary task uses different modalities than the driving task. Thus, hands-free and handheld phone conversations both had a negative influence on driving performance, compared to a single-task condition. This means that even using different modalities, MWL still rises with higher task demand~\cite{Caird2018DoesStudies}. Similar effects could also be detected when interacting with an in-vehicle interface while driving~\cite{Reyes2008EffectsPerformance}.
Traditionally, MWL can be estimated in three ways: using performance measures on the primary task, using self-reported subjective measures, or using psychophysiological measurements such as electroencephalography (EEG), electrocardiogram (ECG), electromyography (EMG), pupillometry, etc.~\cite{Dirican2011PsychophysiologicalInteraction,Moustafa2017AssessmentTechniques,Lohani2019ADriving, Marquart2015ReviewWorkload,tao2019systematic}. However, subjective measures are hard to obtain in real time as the driver's tasks must be interrupted to report his mental state.
Additionally, most of the previous work studying MWL in the driving context use the n-back paradigm as the secondary task~\cite{Solovey2014classifying,mehler2009impact,Mehler2012SensitivityGroups,ross2014investigating,Barua2020TowardsClassification}. The 1,2,3-back versions of the task (the 0-back version is designed for vigilance and does not induce mental workload~\cite{king2015neural}) induce continuous task demand on the working memory, like monitoring, updating information, and rule-based task decisions~\cite{owen2005n}. 
As the performance of the primary task in dual task conditions is modulated by the modality of the secondary task~\cite{Caird2018DoesStudies,wickens2008multiple}, we induce MWL by a secondary task with no attention interference on the driving (i.e., only MWL interference) through an auditory n-back task version.

\subsection {Perceptual Load (PL)}

Another factor affecting task demands is perceptual load. Lavie's PL theory ~\cite{lavie1995perceptual, lavie2005distracted, lavie2004load} states that perceptual capacity is a limited resource that is always involuntarily used. Thus, PL is the extent to which a task consumes available capacity. When perceptual capacity is exhausted due to a high PL, no additional task-irrelevant stimuli are processed, and an inevitable early selective process is triggered~\cite{sperling1960information, treisman1969strategies}. When perceptual capacity is not exhausted, the remaining perceptual resources spill over to task-irrelevant stimuli. This leads to a late selective process that requires available cognitive resources~\cite{deutsch1967comments}.
When cognitive load is kept constant, a higher PL leads to less interference of irrelevant stimuli due to early processing, while a lower PL is related to a higher interference of stimuli. For the driving context, a higher PL would lead to a reduced awareness of unexpected stimuli (e.g., a deer crossing the street) or induce inattentional blindness as well as inattentional deafness~\cite{macdonald2011visual, simons1999gorillas}, possibly leading to a safety hazard in driving.
Despite the load theory being relevant to different processes while driving, there is a lack of applied evidence for the model~\cite{murphy2016twenty}, specifically differentiating between PL and MWL. Murphy \& Greene~\cite{murphy2017load} manipulated PL separately from MWL via a visual search task and found significantly increased inattentional blindness and deafness in the high PL condition and higher reaction time to hazards, thus demonstrating that PL is an important factor in driving performance.
Similarly, we implement a visual search task with variable difficulty. However, the visual search task is implemented on an in-vehicle HMI by varying the set size of stimuli to measure the effect of limiting available information, which is proven to be a factor influencing PL~\cite{lavie2005distracted}. Furthermore, the visual search task mimics the task of navigating an in-vehicle HMI, providing ecological validity.
As far as we know, there is also a lack of studies aiming to connect PL to psychophysiological measurements. Chen \& Epps~\cite{chen2014using} attempted to separate PL and MWL as two distinct sources of task difficulty and found a significantly higher variation in pupil dilation for high PL, while Oliva~\cite{oliva2019pupil} focused on the selection mechanisms of the load theory and found larger peak pupil dilation for the high PL condition.

\subsection{HMI Design in Automotive Domain}

Car manufacturers are continuously adding features to modern cars moving towards smart autonomous vehicles and ease of driving~\cite{Ayoub2019FromAutoUI}. 
However, these features add further complexity to the driving environment~\cite{Solovey2014classifying,BENNAKHI2016Ambient,lee2005driving}. Thus, recent research advancement is oriented towards more user-centered design approaches for in-vehicle interfaces in order to alleviate the mental effort accompanying these added features~\cite{Kern2009DesignInterfaces,Nestler2009CommonUser-interfaces,Olaverri-Monreal2014CapturingExpectations,Gibson2016User-centeredConsoles,Carsten2019HowSolutions}.
Consequently, multiple designs emerged for seamless non-intrusive in-vehicle interfaces~\cite{Korber2013UserContext,Bryant2014DrivingDesign,Sterkenburg2017TowardsStudy,Braun2017ASystems,Francois2017AutomotivePerspectives,Kindelsberger2018DesigningHMI,Ulahannan2020UserVehicle,Gao2020UserCar,Gomaa2020,Aftab2020,gomaa2021ml}.
While these previous approaches and studies focus on enhancing user experience and reducing drivers' MWL through offline pre-design feedback (e.g., gathering users' requirements and designing a universal semi-customizable interface), others focus on real-time (and semi-real-time) approaches to obtain user feedback and adapt the interface based on the user's MWL and stress levels. Hence, a method for estimating MWL through psychophysiological measures~\cite{fridman2018cognitive,Lohani2019ADriving,Cantin2009MentalComplexity,Verwey2000On-lineMeasures,Marquart2015ReviewWorkload,Islam2020ALearning} and PL is needed~\cite{murphy2017load}. 
Researchers~\cite{Mehler2012SensitivityGroups,Solovey2014classifying, tao2019systematic} classified MWL in field studies using two types of psychophysiological dimensions: Skin Conductance Level (SCL) and Heart Rate (HR). However, while a field study provides better external validity, it imposes a safety restriction that limited these studies. They mostly used a simple driving task to avoid accidents, which meant that the primary driving task induced minimal mental load. 
On the other hand, Barua et al.~\cite{Barua2020TowardsClassification} classify MWL using both psychophysiological methods and driving performance in a simulated environment. They were able to introduce a high mentally loading task in the simulation environment and assess the MWL based on the driving performance.
Unlike these previous methods, this paper focuses on both MWL and PL estimation using psychophysiological dimensions. 

\textbf{Our contributions can be summarized as follows}: 1) We propose an MWL and PL estimation approach assessed on a dual task scenario involving a non-trivial driving task. This MWL and PL estimation approach can be used for adapting interfaces, triggering warnings, or managing the transfer of control of semi-autonomous vehicles. 
2) We select psychophysiological dimensions that can be realistically obtained in a non-intrusive way for a real driving scenario (e.g., with a non-wearable eye-tracker and heart rate sensors that currently exist in modern vehicles~\cite{2019BMW2019,2020Toyota2020,2021Mercedes-Benz2021}).
3) We provide a novel multimodal open-source framework with an anonymized dataset for MWL and PL classification in a dual task driving scenario.

\subsection{Empirical Hypotheses}
In this work, we propose an end-to-end predictive ML-based classification framework for MWL and PL estimation. Furthermore, we evaluate the sensitivity of psychophysiological measures in the detection of elevated levels of MWL and PL, paving the way towards adaptive in-vehicle HMIs reactive to momentary changes in MWL and PL. We formulated four hypotheses for this work as follows. 


\begin{itemize}
    \item \textbf{H1}: Correlations between measurements from a similar physiological channel (e.g., heart data) should be higher than correlations between different physiological channels (e.g., heart data and pupillometry)~\cite{Matthews2015TheDivergent}. Previous studies show that convergent validity is higher than discriminant validity for selected MWL measurements~\cite{Matthews2015TheDivergent}. However, since these measurements reflect the same latent construct, we should find a significant inter-correlation. 
    \item \textbf{H2}: Psychophysiological measures can distinguish different n-back task levels while driving in a lane-change task, reflecting a difference in MWL levels~\cite{mehler2009impact}. This distinction between MWL levels should be reflected in both statistical analysis and the predictive ML classifier that would be deployed in the vehicle's system.
    \item \textbf{H3}: Similar to~\cite{ross2014investigating}, MWL levels are distinguishable by the driving performance during the lane-change task.
    \item \textbf{H4}: Similar to the previous hypotheses, PL induced by visual search task affects both psychophysiological measurements and driving performance in the lane-changing task~\cite{chen2014using, oliva2019pupil}. Similar to MWL, the distinction between PL levels should be reflected in both statistical analysis and the predictive ML classifier.
\end{itemize}



\section{Method}

\subsection{Participants}

In total, $49$ participants were recruited for the study\footnote{The number of participants needed for the study was calculated with G*Power~\cite{faul2009statistical}, using an effect size of $f = 0.229$ from~\cite{Matthews2015TheDivergent}. It estimated $n = 45$.}.  
Four participants were excluded for the following reasons: two due to recording errors, one due to a cardiac disease, and one due to incorrect task execution. 
The remaining $45$ participants ($42.2 \%$ female, $2.2\%$ non-binary) with a mean age of 30.07 years ($SD = 10.38$) completed the entire driving route while doing a secondary task. The majority of participants were of German nationality ($62.2 \%$). $48.9 \%$ had a high school degree, $24.4 \%$ had a bachelor's degree, and $26.7 \%$ had a master's degree.
Regarding handedness, $88.9 \%$ of participants were right-handed. 
As for visual acuity, $37.8 \%$ had corrected eyesight in the form of contact lenses only for compatibility with the eye tracker glasses. 
Concerning participants' driving experience, participants had their driver's license on average for $10.20$ years ($SD = 8.74$).  Most participants reported driving often ($13.3 \%$ daily, $44.4 \%$ several times per week, $17.8 \%$ several times per month), and $71.1 \%$ reported previous experience with automatic transmission cars, while only $37.8 \%$ reported previous experience with a simulator.


\begin{figure}[t]
     \centering
    \begin{subfigure}[t]{0.43\textwidth}
         \centering
         \includegraphics[width=6cm,height=4cm,keepaspectratio]{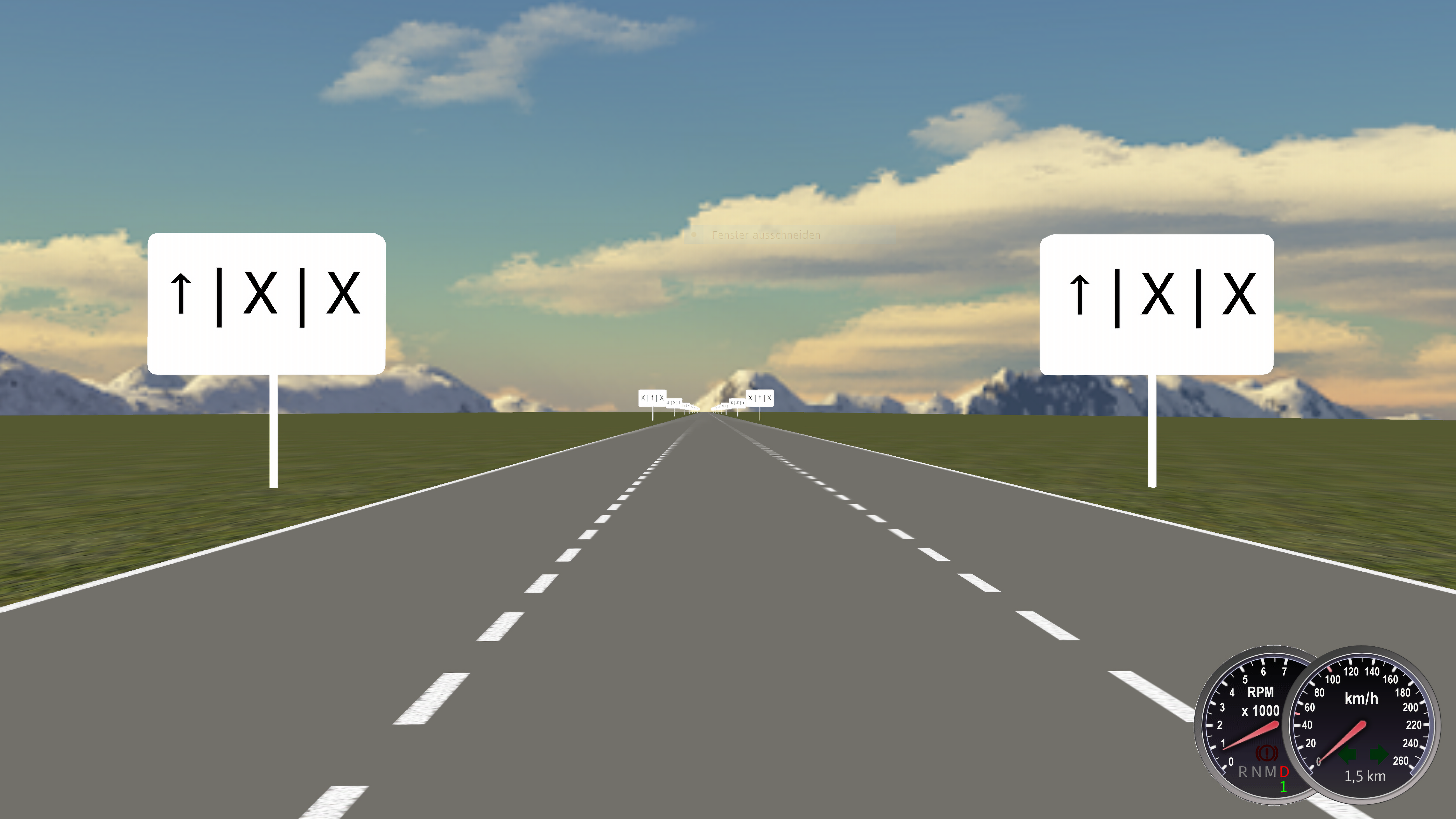}
         \caption{}
         \label{fig:lanechanging}
     \end{subfigure}
    \begin{subfigure}[t]{0.43\textwidth}
         \centering
         \includegraphics[width=6cm,height=4cm,keepaspectratio]{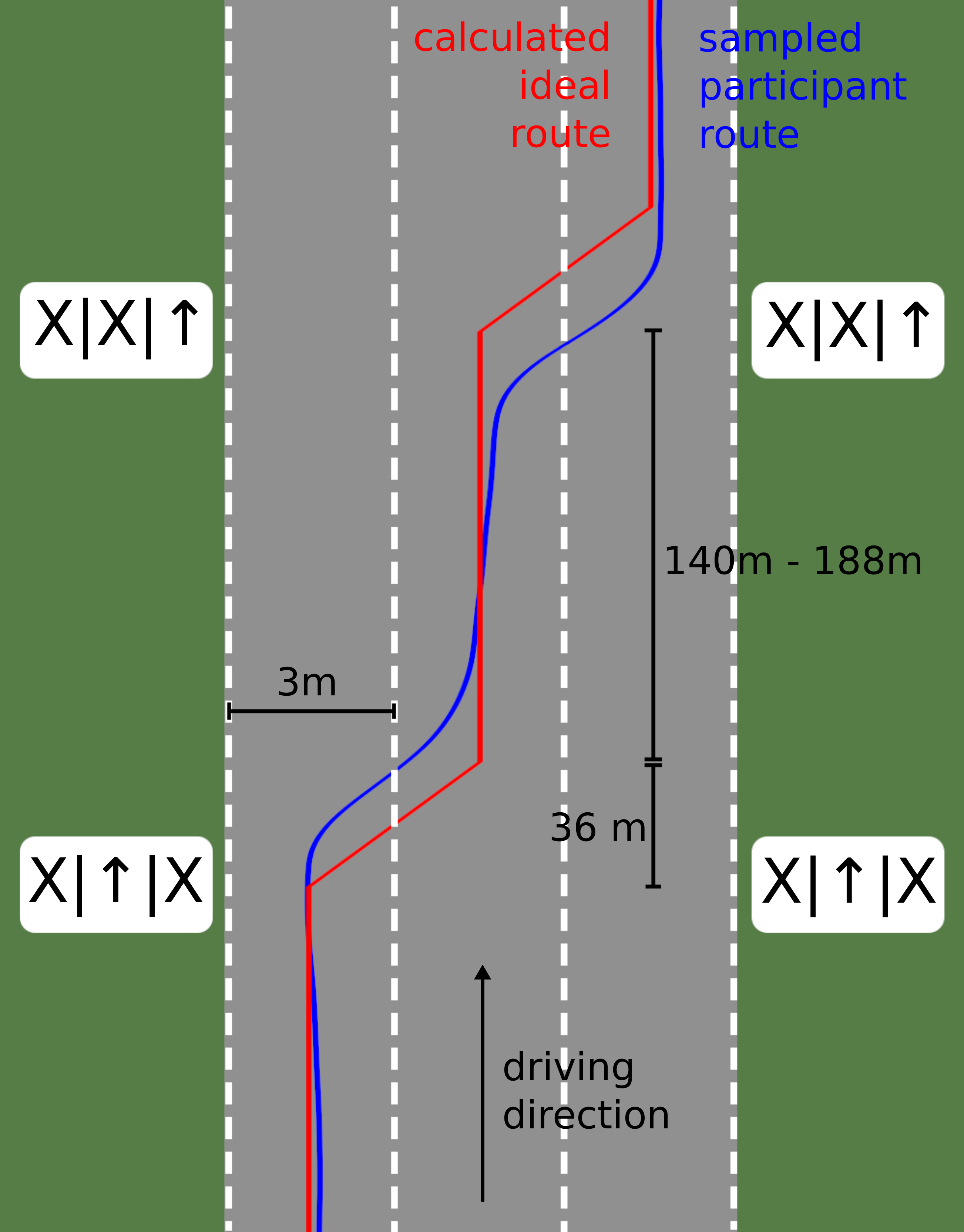}
         \caption{}
         \label{fig:primarytaskmetrics}
     \end{subfigure}
     \caption{Our lane-changing task. (a) OpenDS driving simulator. (b) Schema (not to scale) of driving performance evaluation as in~\cite{Mattes2003LaneTask}.}
     \Description{Figure of the lane-changing task. The left image shows a screenshot from inside OpenDS. The right side shows an overview of the lane-changing task. The ideal path is depicted as red line, while the path taken by the driver as a blue line.}
\end{figure}

\subsection{Design}

We designed a within-subject dual task counterbalanced driving experiment in a simulated study~\cite{math2013opends}. The participants completed a primary lane-changing task (LCT) on a straight three-lane road along each of the secondary tasks. For the secondary task, the participants performed both a visual search task (PL-correlated stimuli) and an n-back task (MWL-correlated stimuli) separately.
Each of the two secondary tasks was conducted in three different difficulty levels (easy, medium, and hard), resulting in a $2x3$ within-subject factor design. The experiment was piloted to mitigate design issues.
For the primary driving task, our LCT was adapted from~\cite{Mattes2003LaneTask}, a procedure commonly employed in driving simulator studies~\cite{Reich2014ToEnvironments,Huemer2010Alcohol-relatedTask,Burnett2013HowMethods}. The target lane was marked with an upward arrow ($\uparrow$), while the other two lanes were marked with an (X) symbol.
The lane-change task was divided into 6 road segments: one segment per difficulty level per secondary task, and it lasted between 120 to 160 seconds based on the driving speed with a 90-second pause between segments to allow the workload to return to baseline. There were no vehicular traffic or environmental changes to avoid confounding factors.
\begin{figure}[b]
     \centering
     \begin{subfigure}{0.3\linewidth}
         \centering
         \includegraphics[width=\textwidth]{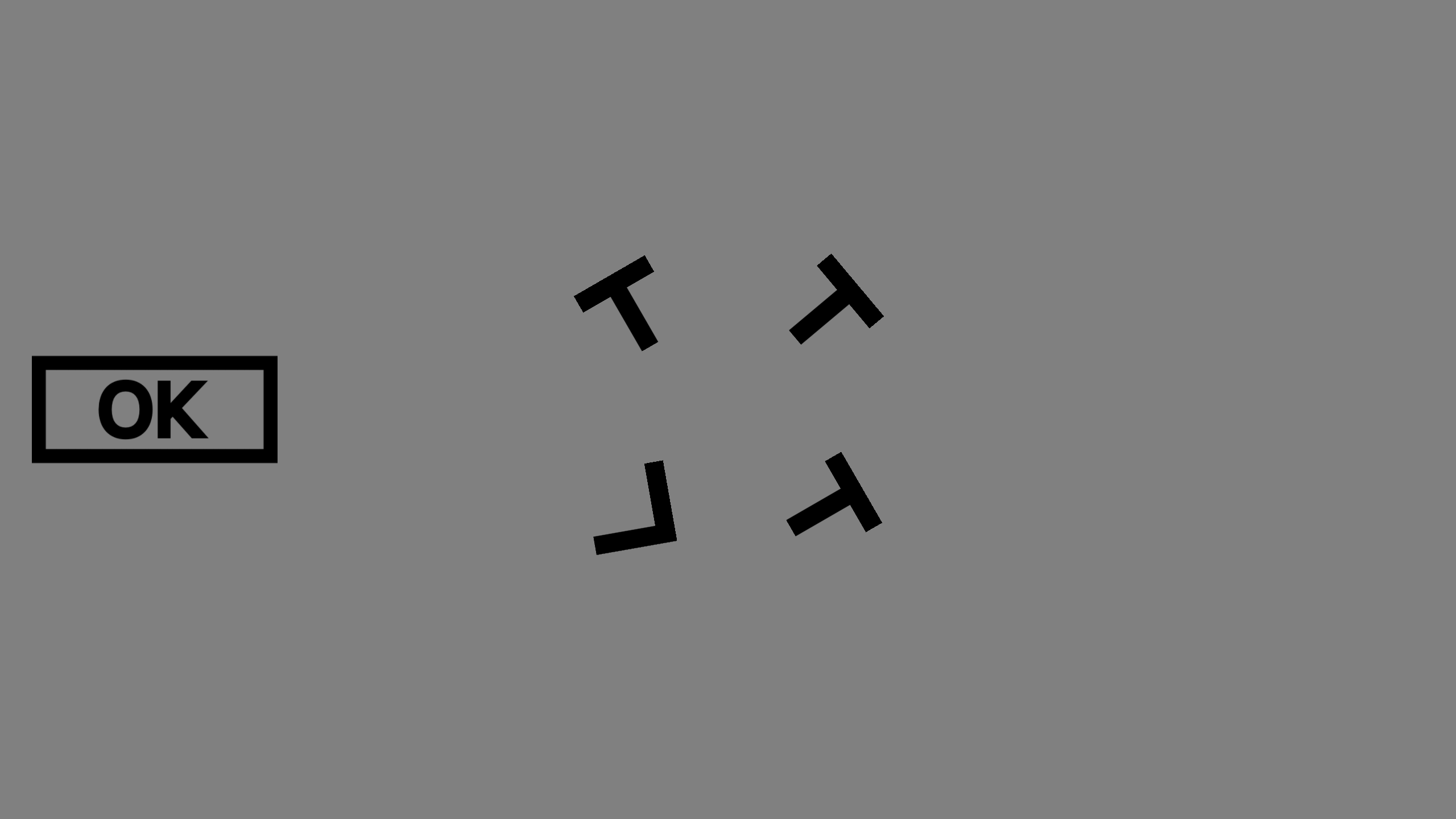}
     \end{subfigure}
     \begin{subfigure}{0.3\linewidth}
         \centering
         \includegraphics[width=\textwidth]{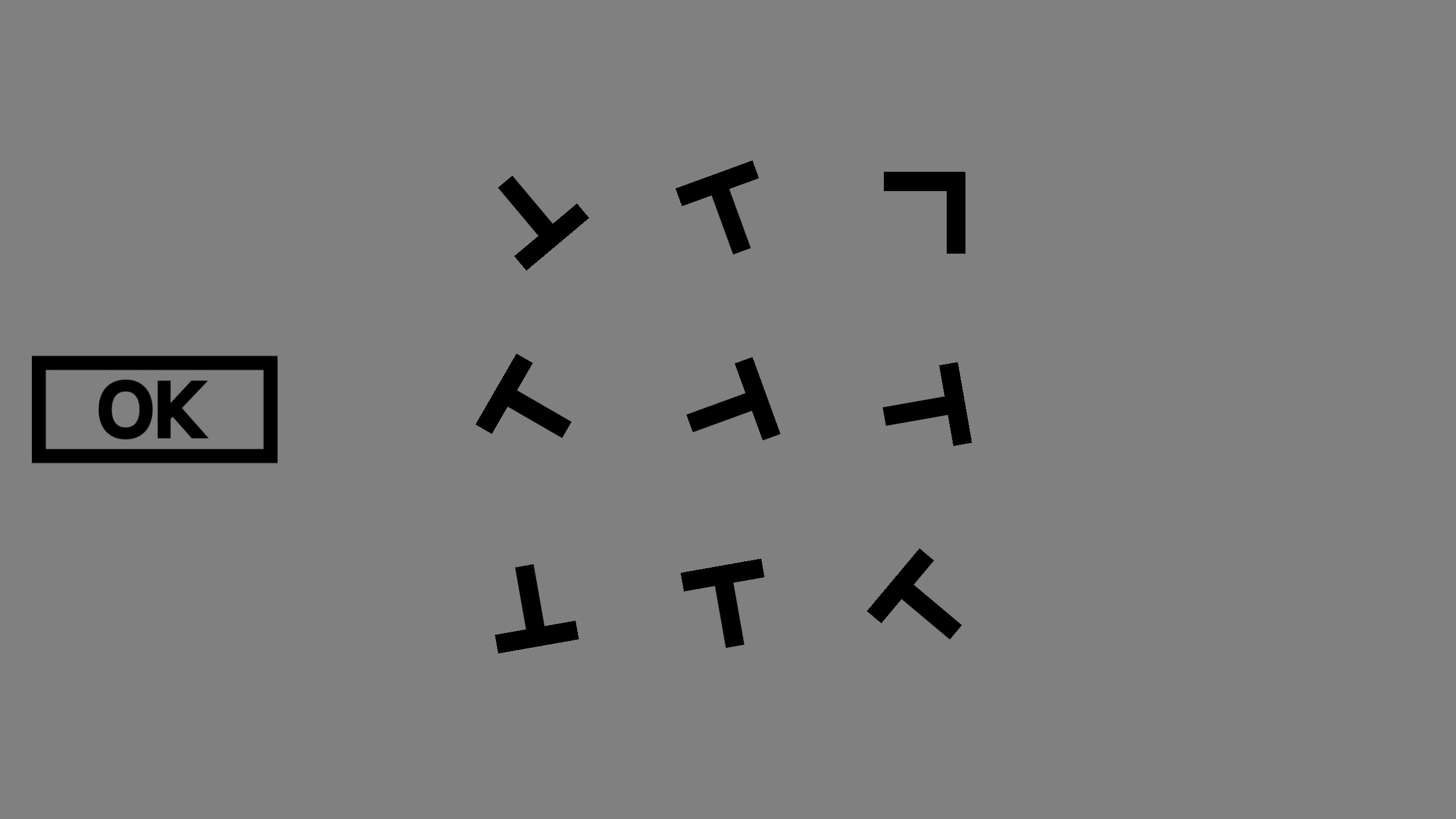}
     \end{subfigure}
          \begin{subfigure}{0.3\linewidth}
         \centering
         \includegraphics[width=\textwidth]{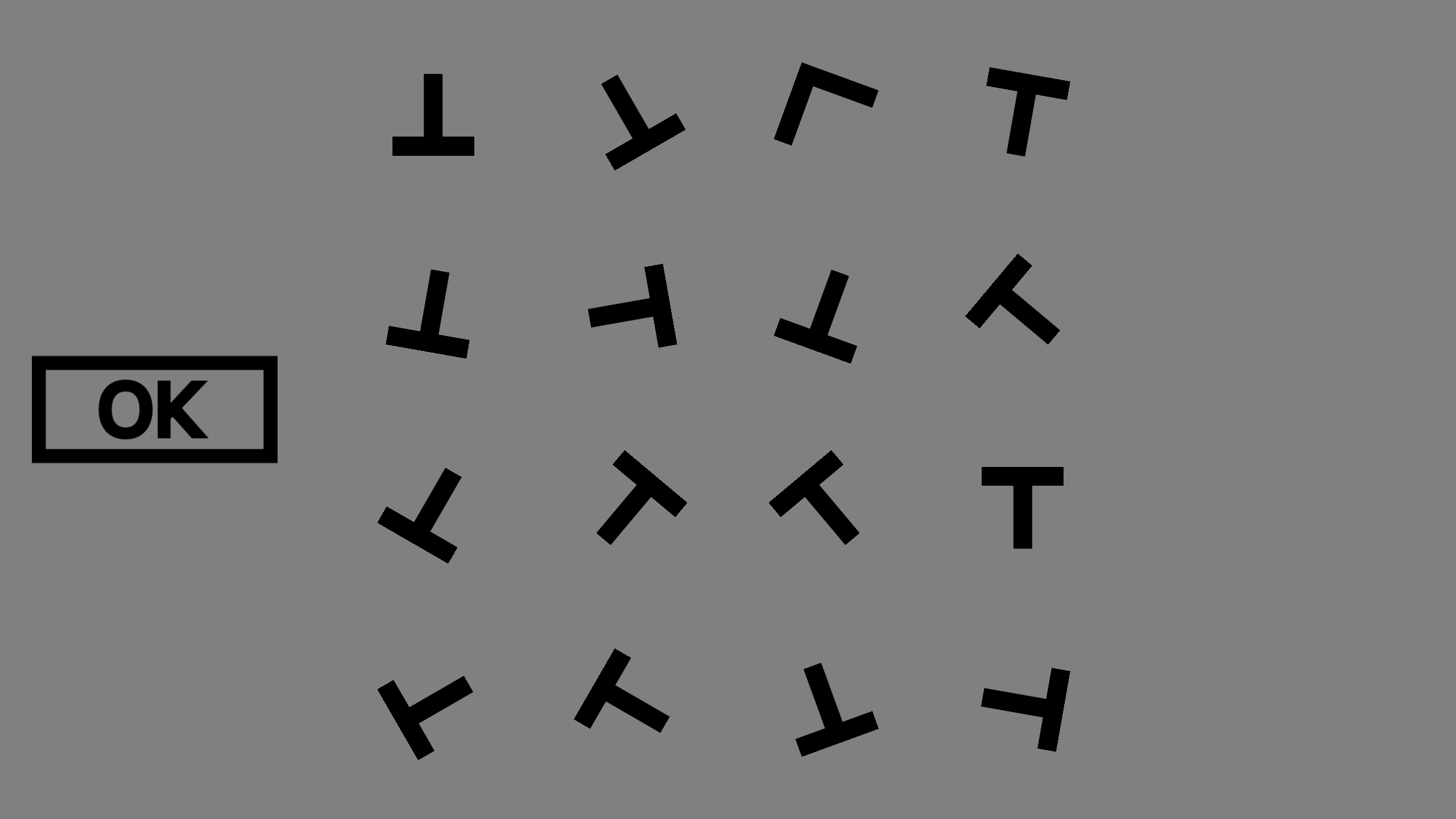}
     \end{subfigure}
     \caption{The stimulus array for the visual search task per difficulty level. From left to right are the $2x2$, $3x3$, and $4x4$ matrices.}
     \Description{The image shows the three stimulus arrays for the visual search task. The three matrices from left to right are of size 2 times 2, 3 times 3, and 4 times 4. The task is to find the L among the distractors T.}
     \label{fig:visualsearch}
\end{figure}
For the visual secondary task, a visual search task~\cite{lavie2005distracted} was presented on a tablet on the right side of the steering wheel. This task simulates the distraction of interacting with the in-vehicle infotainment systems. The task difficulty was manipulated by the number of items on the display (i.e., set size). The levels consisted of a $2x2$, $3x3$, and $4x4$ matrix of items (see~\autoref{fig:visualsearch}). Each level had 40 stimulus arrays. Each stimulus array was presented for $2000$ms with a $1000$ms pause in between. During this stimulus, the participant had to identify the existence of an L-shaped target item among T-shaped distracting items and response by pressing an ``OK'' button only (see~\autoref{fig:visualsearch}). All the items were randomly rotated between $0\degree$ and $360\degree$. L-shaped targets were present in $50\%$ of the trials. 
As for the auditory secondary task, an n-back task was utilized where participants had to listen to numbers and respond when the current number matched a previously presented number. This task simulates the distraction of talking on a phone or to a passenger. Numbers from zero to nine were used as stimuli and presented in German through automated speech. Each number was presented for the three difficulty levels consisting of a $1$-back, $2$-back, and $3$-back task. The task was to listen to each number and to respond by saying ``Yes'' in German (i.e., ``Ja'') when the currently presented number was the same as the number presented one, two, or three positions before it, respectively. The numbers were presented in total 40 times, as in the visual search task; however, targets were present in $25\%$ of the cases. 
Responses were coded manually and were verified with an inter-rater reliability analysis~\cite{ranganathan2017common}. Random samples were assessed by a second rater from a video recording. An intra-class correlation (ICC) was calculated with SPSS using a two-way mixed model yielding a value of $r = .96$ $(F (17, 17) = 33.29$, $p < .001)$ constituting excellent reliability.


\subsection{Apparatus and Procedure}

\begin{figure}[t]
	\begin{center}
		\includegraphics[width=\linewidth]{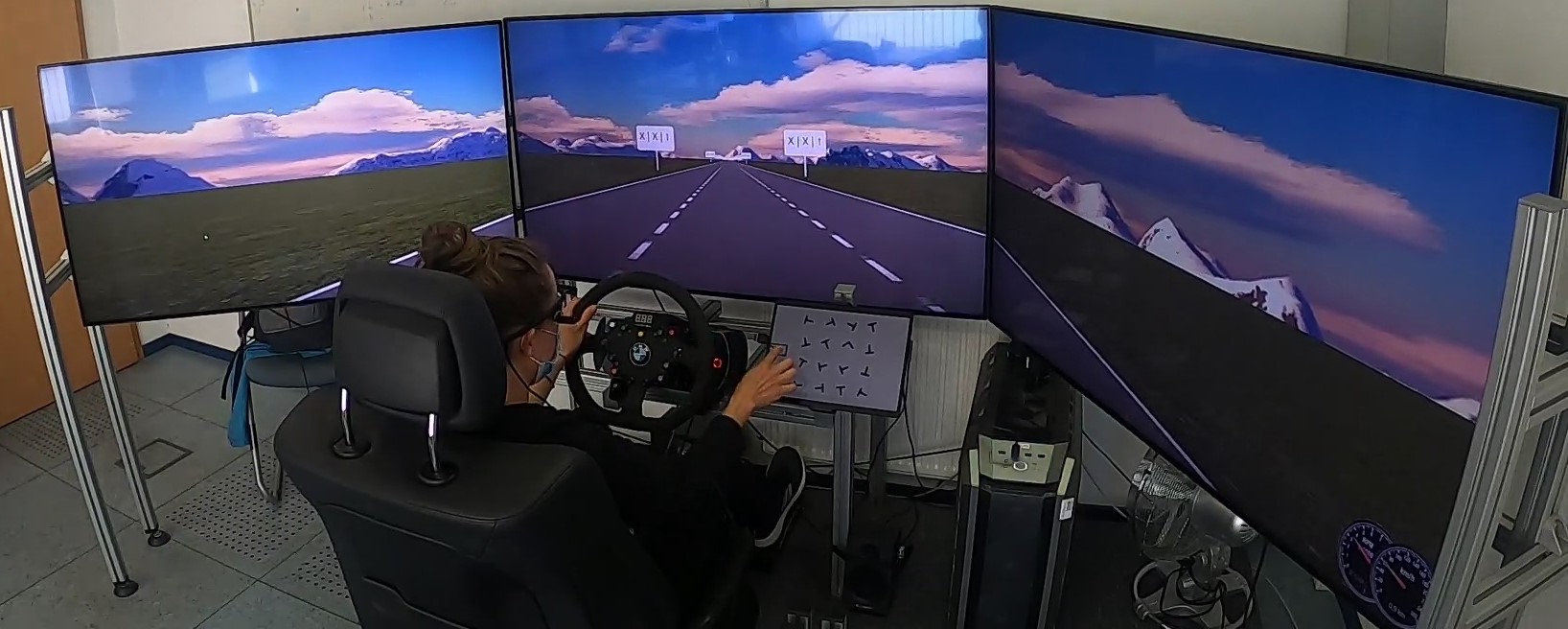}
	\end{center}
	\caption{Driving simulator setup overview showing the lane-changing task and the secondary visual search task.}
	\Description{The image shows the setup of the simulator with three screens and a tablet right of the steering wheel, simulating the infotainment system.}
	\label{fig:setup2}
\end{figure}

The driving simulator setup consisted of a driver's seat surrounded by three $55$-inch LCD screens, two were rotated at a $45$-degree angle towards the driver to allow for a wider field of view, as in~\autoref{fig:setup2}. The simulator vehicle was on automatic transmission and controlled via a steering wheel, and gas and brake pedals. Heart rate data was collected through a commercial ECG-based sensor attached to the chest (Polar H10)
and eye data was collected by a head-mounted Pupil Labs Core eye-tracker~\cite{Kassner2014pupil}.
Participants were introduced to the sensors and completed a short training in the LCT and secondary tasks. Then, they completed the designed dual task procedure while instructed to prioritize the secondary tasks. 
Both the difficulty levels of each secondary task as well as the order of the tasks themselves were counterbalanced. Finally, they filled out a post-experiment questionnaire and received monetary compensation.

\begin{figure*}[t]
	\begin{center}
		\includegraphics[width=\linewidth]{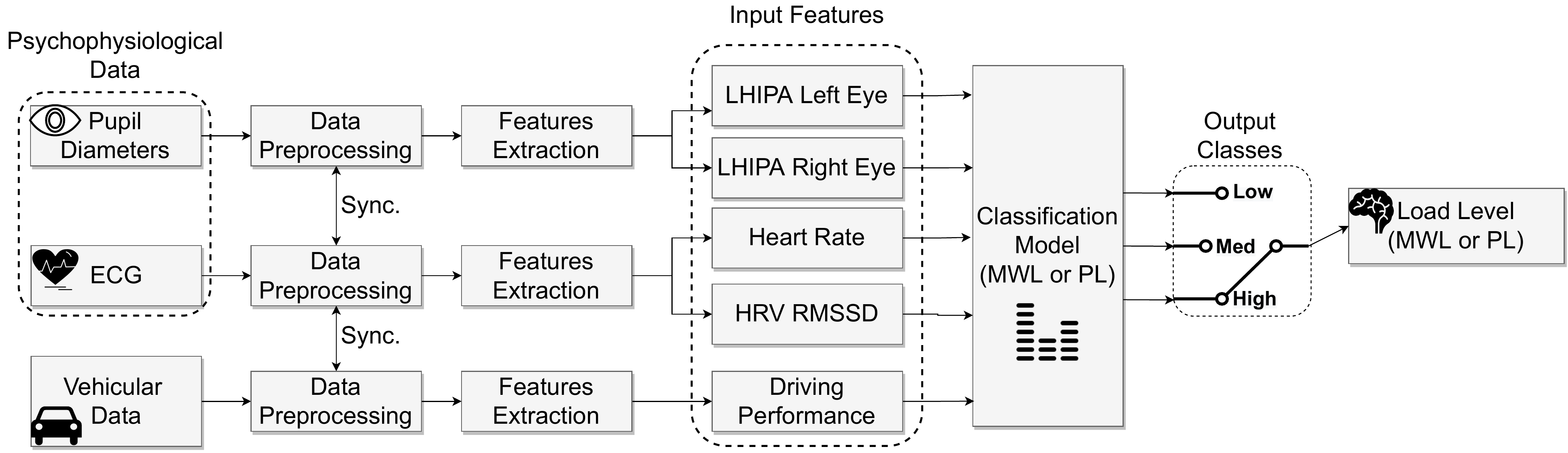}
	\end{center}
	\caption{System architecture showing our minimally intrusive framework for MWL and PL estimation separately.}
	\Description{System Architecture. The left side shows the physchophysiological data capturing. This data is then preprocessed and the features extracted. This results in the input features for the machine learning classification model. The classification model then predicts one of the load levels (either PL or MWL) as classes.}
	\label{fig:systemoverview}
\end{figure*}

\subsection{Data Preprocessing and Measurements}

In this work, we rely on two non-intrusive psychophysiological (ECG and pupillometry) measurements, as well as primary and secondary task performance, to estimate the driver's MWL and PL. These are described as follows.

\paragraph{Psychophysiological Measurements}

An electrocardiogram (ECG) measures the electrical signals from a person's heart to determine the heart rate (HR) in beats per minute. 
Heart rate variability (HRV)~\cite{Murai2004TheVariability} is defined as the variation of the peak-to-peak time interval (RR interval) between two consecutive heartbeats.
According to several previous studies, average HR and HRV can be used to estimate MWL levels~\cite{Delliaux2019MentalDynamics} where an increase in MWL level would increase HR values while decreasing the HRV values. 
Multiple features can be extracted from both HR and HRV over the time interval, indicating the MWL and PL level. 
For HR, we used average HR, minimum, maximum, and standard deviation. As for HRV, there are several features divided into three categories: time domain, frequency domain, and non-linear features. However, we used only the time-domain-related root mean square of the successive differences between RR intervals (RMSSD) feature, which is the most used feature for mental load level estimation with HRV. 
Pupillometry is defined as the study of pupil size (i.e., diameter) and reaction. According to~\cite{Beatty1982Task-evokedResources.,beatty2000pupillary,van2018pupil,oliva2019pupil}, it is possible to estimate MWL and PL levels using pupil diameter. 
However, pupil diameter is not lighting invariant, which makes it unsuitable for a generic light-independent setting~\cite{van2018pupil}. 
Hence, researchers attempted different preprocessing steps that would result in a light-invariant pupil-dependent method for mental load estimation, such as the index of cognitive activity (ICA)~\cite{Marshall2002TheWorkload}, the index of pupillary activity (IPA)~\cite{Duchowski2018TheOscillation}, and the low/high index of pupillary activity (LHIPA)~\cite{Duchowski2020TheActivity}. 
While ICA is the most common method in previous studies, it is a patented approach. 
Therefore, Duchowski et al. developed the IPA and LHIPA as an open-source alternative for ICA. The IPA algorithm is constrained to a fixed-head setting only, which does not apply to a generic use-case such as driving, while LHIPA is an enhancement to the IPA algorithm with fewer constraints and more applicability. In our study, we use LHIPA as an input feature to our algorithm. We calculate it for each eye separately based on their respective pupil diameter variations.

\paragraph{Primary and Secondary Tasks Measurements}
For the driving performance, the participant's position in the driving task was compared to a pre-calculated ideal path in terms of lateral deviation in meters with a 33 Hz sampling rate. A standardized procedure (ISO 26022:2010) commonly used to access the LCT driving performance~\cite{Mattes2003LaneTask,Huemer2010Alcohol-relatedTask,Burnett2013HowMethods,Chong2014AuditoryLoad,Burns2005MeasuringLCT,Pitts2012EvaluatingTest}.  The results were sorted into their corresponding tasks and different metrics including average deviation, median, minimum and maximum, and standard deviation. 
The ideal path was defined to be the center of the lane where the driver should drive except when close to lane changes. Pilot runs showed that drivers needed approximately 36 meters to complete the lane change when driving undisturbed at 60 km/h, which was considered when defining the ideal path.
~\autoref{fig:primarytaskmetrics} highlights the structure of the ideal route and its comparison to the actual driving. Since the vehicle was on automatic transmission, further analysis of gear changing, clutch pressure, and other vehicle-related measurements is not applicable.
In the visual search task, performance was measured as the reaction time between stimulus onset and the pressing of the ``OK'' button. As for the n-back task, performance was calculated by subtracting false positives 
from correct responses 
and dividing the result by the total number of possible correct responses. 

\subsection{Classification Framework}

The classification task aims to differentiate between different MWL levels for the performing the n-back task while driving and different PL levels for performing the visual search task while driving.~\autoref{fig:systemoverview} shows an overview of the MWL or PL level classification approach. It starts with the data collection and preprocessing stage (discussed earlier), followed by the input features extraction stage, and ends with a model training and prediction stage to determine the MWL or PL level (based on the load-inducing task). Features extraction and the ML models are described as follows.

\subsubsection{Input Features and Output Classes}

To classify the different MWL levels based on the acquired measurements, we utilize machine learning (ML) techniques to learn from the features generated from these measurements and create a prediction model for real-time inference during deployment.~\autoref{tab:input_features} shows the generated possible input features per output class. In total, our model has eight possible input features calculated per output class. The classification model uses these input features to classify between two pairs of three output classes (i.e., easy, medium, and hard) for each n-back task (inducing MWL changes) and visual search task (inducing PL changes).

\begin{table}[b]
\centering
\renewcommand{\arraystretch}{1.2}
\caption{Input features generated for the machine learning models.}
\Description{Table containing the input features for the machine learning models. These features come for one from the ECG like the mean, minimum and max heart rate as well as the HRV. Other features come from pupillometry, like the low/high index of pupillary activity for both eyes. And finally, driving performance. }
\resizebox{\linewidth}{!}{
\begin{tabular}{l p{0.4\linewidth} c}
\hline
Data Type                   & Description of Input Features                                                    & Number of Input Features \\ \hline
ECG          & Mean, minimum, maximum, standard deviation of heart rate, and HRV-RMSSD & 5                             \\
Pupillometry & Left and right eyes' low/high index of pupillary activity (LHIPA)  & 2                            \\
Driving Performance         & Average deviation during lane switching  & 1                                              \\ \hline
\end{tabular}}
\label{tab:input_features}
\end{table}

\begin{table*}[t]
\centering
\renewcommand{\arraystretch}{1.1}
\caption{Descriptive analysis (Mean$\pm$Std) of the psychophysiological measurements and driving performance per condition.}
\Description{The table shows the descriptive analysis of the psychophysiological measurements  and the driving performance per secondary task and per condition. To note is that the values don't seem significantly different in the visual search task in contrast to the n-back task, where they seem to be more different.}
\resizebox{0.85\linewidth}{!}{
\begin{tabular}{lccccccc}
\hline
                      & \multicolumn{3}{c}{N-Back Difficulty}                    & \multicolumn{1}{c}{} & \multicolumn{3}{c}{Visual Search Difficulty}             \\ \cline{2-4} \cline{6-8} 
                      & Easy          & Medium        & Hard          &                      & Easy          & Medium        & Hard          \\ \hline
Heart Rate            & $77.49\pm12.60$ & $82.54\pm14.17$ & $82.97\pm14.78$ &                      & $77.29\pm11.81$ & $78.58\pm12.00$ & $78.63\pm12.7$  \\
HRV-RMSSD & $37.79\pm19.50$ & $31.86\pm15.99$ & $30.34\pm14.42$ &                      & $38.37\pm18.53$ & $36.52\pm17.20$ & $37.70\pm19.36$ \\
                      &               &               &               &                      &               &               &               \\
LHIPA Right Eye           & $2.38\pm0.50$ & $2.28\pm0.31$ & $2.29\pm0.43$ &                      & $2.46\pm0.58$ & $2.39\pm0.56$ & $2.44\pm0.58$   \\
LHIPA Left Eye           & $2.37\pm0.51$ & $2.34\pm0.34$ & $2.29\pm0.30$ &                      & $2.3\pm0.22$ & $2.3\pm0.24$ & $2.30\pm0.24$   \\
                      &               &               &               &                      &               &               &               \\
Driving Performance   & $0.15\pm0.08$ & $0.21\pm0.13$ & $0.23\pm0.16$ &                      & $0.24\pm0.11$ & $0.24\pm0.12$ & $0.27\pm0.12$   \\ \hline
\end{tabular}}
\label{tab:Deskr_Final}
\end{table*}

\subsubsection{Dataset Split}

The entire dataset covers 45 participants and is split into training, validation, and test sets using an adaptation of the nested cross-validation~\cite{Browne2000} approach with 5-fold cross-validation in the outer loop and a one-third holdout test set in the inner loop. Note that all the splits are done on the participant level and not on the data point level. This way, no data from the same participant is used in training, validation, or testing for the same model. This is to ensure external validity and model generalization while avoiding data leakage~\cite{kaufman2012leakage}.

\subsubsection{Machine Learning Models}

For our machine learning architecture, we utilize Fuerer et al.'s~\cite{fuerer2015autosklearn} auto-sklearn python library for classification (based on scikit-learn \cite{scikit-learn}) that automates finding the optimal machine learning model and its corresponding hyperparameters. This approach automatically searches for the optimum machine learning model using Bayesian optimization methods in a structured search space consisting of the combination of 15 classifiers, 14 feature preprocessing methods, and 4 data preprocessing methods. It also utilizes an ensemble approach that combines several optimum machine learning models to enhance prediction accuracy~\cite{Caruana2004ensemble}. In this method, we compare an ensemble approach to one of the top 3 best-performing classifiers selected by the Bayesian optimization process. The best-performing classifiers for our dataset were the k-nearest neighbors (KNN)~\cite{goldberger2004neighbourhood}, the linear discriminant analysis (LDA) classifier~\cite{ledoit2004honey}, and AdaBoost~\cite{freund1997decision}.

\section{Results}

Results are divided into two sections. Section~\ref{sec:stats_results} shows the statistical results, 
while section~\ref{sec:ML_results} shows the results of the multimodal machine learning based MWL and PL classification models.

\subsection{Statistical Results}
\label{sec:stats_results}

For the main statistical analysis, statistical significance is determined using a p-value below $0.05$ unless otherwise mentioned. Extreme outliers were excluded; thus, the number of participants varies between analyses.


\subsubsection{Descriptive Analysis}

The psychophysiological data was evaluated for each condition of each secondary task and averaged across all conditions. The subjects showed a mean HR of $79.58$ bpm ($SD=12.57$), a mean HRV-RMSSD of $35.29$ milliseconds ($SD=16.72$), and a mean LHIPA value of $2.37$ ($SD=0.25$) for the right eye and $2.33$ ($SD=0.16$) for the left eye. Mean driving performance was $0.22$ ($SD = 0.24$). Further mean and standard deviation details are shown in~\autoref{tab:Deskr_Final}.


\subsubsection{Correlations and Reliability}
To investigate \textbf{H1} (convergent and discriminant validity), we calculated Pearson's correlation for the averaged psychophysiological measurements and additionally for all conditions. Only HRV-RMSSD and HR were significantly correlated ($r(44)=-.53; p<.001$). The higher the HR, the lower the HRV-RMSSD. The average values of the LHIPA did not reveal any significant relationship with any of the other dimensions (all $p$ values $>.177$). 
The same pattern was found when comparing each dimension for each secondary task per difficulty level. HR and HRV-RMSSD were without exception highly inter-correlated, but the values for LHIPA showed almost non-significant interactions, whether with each other or with any of the heart rate measurements. The detailed correlations for each secondary task per difficulty level and per measurement can be found in~\autoref{sec:appendix_corr}.

For reliability analysis, we calculated Cronbach's alpha to analyze the internal consistency of the data \cite{wentz2012dimitrov}. We performed an analysis for each of the five dimensions for the six conditions. HR and HRV-RMSSD yielded significant internal consistencies (HR: $\alpha= .98$; HRV-RMSSD: $\alpha= .98$). As expected by the correlations, the reliability for LHIPA was unacceptable (right eye: $\alpha= .36$; left eye: $\alpha= .31$). Therefore the reliability criteria for LHIPA was not met for \textbf{H2} (MWL distinction) and \textbf{H4} (PL distinction) and had to be excluded from the analyses. The internal consistency of the driving performance showed a Cronbach’s alpha of $\alpha = .33$, which is considered as unacceptable. Consequently, we excluded driving performance for further analysis in the statistical testing \cite{nunnally1994psychometric, bland1997altman}, since the reliability criteria for driving performance was not met for \textbf{H3} and \textbf{H4}.

\begin{table*}[t]
\centering
\renewcommand{\arraystretch}{1.5}
\caption{N-back multi-class average test accuracy and standard deviation (i.e., Mean\%$\pm$Std). The random chance level is 33.33\%.}
\Description{Mean accuracy with standard deviation for the test set in the n-back classification task, using 5 fold cross-validation. The random chance level is 33.33 percent. The heart alone model with ensemble shows the most promising results with $48.9$ plus minus $11.1$ percent.}
\resizebox{0.75\linewidth}{!}{
\begin{tabular}{p{1.2cm}>{\centering\arraybackslash}p{1.2cm}>{\centering\arraybackslash}p{1.8cm}>{\centering\arraybackslash}p{1.8cm}>{\centering\arraybackslash}p{1.8cm}>{\centering\arraybackslash}p{2cm}>{\centering\arraybackslash}p{1.8cm}}
\hline
\multicolumn{2}{c}{}                                   & All Features & Eye \& Drive  & Heart \& Eye & Heart \& Drive  & Heart alone \\ \hline
\multirow{3}{1cm}{Without Ensemble} & LDA              & $43.7\pm7.9$  & $36.3\pm2.8$              & $42.2\pm9.0$ & $43.7\pm6.4$                    & $43.7\pm7.2$  \\ 
                                    & KNN & $42.2\pm4.4$   & $31.9\pm9.0$                & $45.9\pm7.6$  & $43.0\pm9.8$                   & $40.0\pm7.2$  \\ 
                                    & AdaBoost         & $38.5\pm9.8$  & $28.9\pm10.6$                & $43.0\pm10.6$ & $43.0\pm1.8$                  & $38.5\pm11.1$  \\ 
\multicolumn{2}{c}{With Ensemble}                      & $48.9\pm4.9$     & $37.0\pm7.8$               & $42.2\pm5.5$   & $45.9\pm10.1$                   & $48.9\pm11.1$     \\ \hline
\end{tabular}}
\label{tab:machinelearningresultsallclasses}
\end{table*}

\subsubsection{Effects of Mental Workload and Perceptual Load}

To test if the manipulation of the study for MWL and PL was successful, we performed paired t-tests per secondary task. For the n-back task, the performance rate (correct response - false positives / total number of possible correct responses) was used to indicate effects of induced MWL per level. Participants were significantly better at the easy condition ($M = 0.96$; $SD = 0.07$) compared to the medium level ($M = 0.85$; $SD = 0.15$; $t(42) = 4.72$; $p < .001$) and were significantly better at the medium level compared to the hard level ($M = 0.36; SD = 0.21$; $t(41) = 11.86$; $p < .001$).
For the visual search task, we used the reaction time as a suitable indicator of task performance. This was the case because the performance rate was close to 100 \% for most participants in all three levels (easy: $M = 0.99$, $SD = 0.02$; medium: $M = 0.99$, $SD = 0.02$; hard: $M = 0.95$; $SD = 0.04$).
Participants reacted significantly faster in the easy level ($M = 1.29$; $SD = 0.16$) compared to the medium level ($M = 1.44$; $SD = 0.18$; $t(44) = -10.28$; $p < .001$) and significantly faster in the medium level compared to the hard level ($M = 1.75$; $SD = 0.22$; $t(44) = -16.78$; $p < .001$). 
This indicates that the manipulation check was successful, and the levels were sufficiently different.

Additionally, we performed two separate MANOVAs to test the effect of MWL and PL on HR and HRV-RMSSD individually. 
To test \textbf{H2} for differences in MWL levels across various physiological measures, we calculated a MANOVA with the n-back task levels for HR and HRV-RMSSD. The n-back task yielded a highly significant effect for HR and HR-RMSSD ($Wilks-\lambda = .409$; $F (4,40) = 14.45$; $p < .001$; $\eta_p^2 = .591$). Both univariate ANOVAs showed a significant result for HR ($F (1.68,72.28) = 32.19$; $p < .001$; $\eta_p^2 = .428$; Greenhouse-Geisser corrected) and likewise for HRV-RMSSD ($F (1.55,66.48) = 19.50$; $p < .001$; $\eta_p^2 = .312$; Greenhouse-Geisser corrected). 
Helmert contrasts revealed that for HR and HRV-RMSSD, this effect is derived from the difference between the easy level compared to the medium and hard ones combined (HR: $F (1,43) = 47.46$; $p < .001$; $\eta_p^2 = .525$; HRV-RMSSD: $F (1,43) = 24.35$; $p < .001$; $\eta_p^2 = .362$). No significant effect was found between the medium level and hard level, neither for HR ($F (1,43) = 0.68$; $p = .413$; $\eta_p^2 = .016$) nor for HRV-RMSSD ($F (1,43) = 3.19$; $p = .081$; $\eta_p^2 = .069$).
This means that participants showed a lower heart rate in the easy level compared to the medium and hard levels combined, while the HR and HRV-RMSSD remained the same in the medium and hard load levels as seen in~\autoref{tab:Deskr_Final}.
A second MANOVA was performed to test \textbf{H4} and analyze the influence of the different PL levels (set sizes in the visual search task) on the HR and HRV-RMSSD. There was a marginally non-significant effect for HR and HRV-RMSSD ($Wilks-\lambda = .807$; $F (4,41) = 2.45$; $p = .062$; $\eta_p^2 = .193$). There was also a marginally non-significant result in the univariate analysis for HR ($F (1.74,76.67) = 3.27$; $p = .050$; $\eta_p^2 = .069$; Greenhouse-Geisser corrected). Notably, the Helmert contrast for HR revealed that there was a significant effect for the easy level compared to the medium and hard ones in HR($F (1,44) = 10.39$; $p = .002$; $\eta_p^2 = .191$), while the medium level compared to the hard one was not significant ($F (1,44) < 0.01$; $p = .984$; $\eta_p^2 < .001$). There was no univariate effect in HRV-RMSSD for PL ($F (1.77,77.96) = 1.03$; $p = .360$; $\eta_p^2 = .023$).




\subsection{Machine Learning Results}
\label{sec:ML_results}

In this section, we show the results of the machine learning analysis using different learning models. As mentioned earlier, we compare an ensemble-based prediction approach vs. a single-model, and we test them using a 5-fold nested cross-validation approach. This approach was implemented for the two secondary tasks: the n-back task to estimate MWL level and the visual search task to estimate PL level.
\autoref{tab:machinelearningresultsallclasses} highlights the classification results of the n-back task (i.e., MWL classification). It can be seen that using an ensemble model approach is superior to the use of a single model approach for all input feature types. This suggests high complexity in the input features that are difficult to learn using a single model approach for the multi-class classification use case. 
Moreover, it can be seen that the model combining eye features (i.e., LHIPA) and driving performance features alone (without the heart rate features) outputs random chance accuracy. This is in line with the statistical analysis which shows that both eye features and driving performance are unreliable (according to the Cronbach's alpha analysis). This can also be seen when comparing the ``all features'' model with the ``heart data only'' model as both have almost identical accuracy for all the applied learning models.

Moreover, the statistical analysis also showed that only the 1-back task (i.e., low load) is statistically significant with higher load levels. This is also seen in~\autoref{tab:machinelearningresultsallclasses} as the accuracy values are on average only 49\% with a maximum of 63\%. Therefore, a further binary classification model was implemented to classify between the two classes low load and medium load.~\autoref{tab:machinelearningresultsbinary} highlights the results for this binary classification which also aligns with the statistical analysis. It shows that eye features and driving performance features are performing only at a random chance level, while the heart rate data has an average classification accuracy of $72.2\%$, up to $89\%$ for a single fold. Additionally, it can be seen that using ensemble models is comparable to the use of a single model approach. This suggests that the difference in heart rate features is more pronounced for the binary classification approach compared to the multi-class one. 
As for the visual search task (i.e., PL classification), all the machine learning models resulted in random chance accuracy. This result also aligns with the statistical analysis that shows minor statistical significance, although the manipulation check was successful. This would suggest that while the visual search difficulty levels were comparable to each other, they had a small effect on the selected psychophysiological dimensions. Similar to the MWL classification, a binary classification approach was attempted. However, it did not improve the accuracy values compared to the multi-class approach.

\begin{table*}[t]
\centering
\renewcommand{\arraystretch}{1.5}
\caption{N-back binary-class average test accuracy and standard deviation (i.e., Mean\%$\pm$Std). The random chance level is 50\%.}
\Description{Table of mean accuracy and standard deviation for the test set with a 5-fold cross-validation in the n-back binary classification task. Random chance level is 50 percent. The results show the best prediction model to be the AdaBoost using all features with about $73.3$ percent accuracy. However, other models like KNN with all features or the ensemble model with heart rate data only don't lag behind too much.}
\resizebox{0.75\linewidth}{!}{
\begin{tabular}{p{1.2cm}>{\centering\arraybackslash}p{1.2cm}>{\centering\arraybackslash}p{1.8cm}>{\centering\arraybackslash}p{1.8cm}>{\centering\arraybackslash}p{1.8cm}>{\centering\arraybackslash}p{2cm}>{\centering\arraybackslash}p{1.8cm}}
\hline
\multicolumn{2}{c}{}                                   & All Features & Eye \& Drive  & Heart \& Eye & Heart \& Drive  & Heart Alone \\ \hline 
\multirow{3}{1cm}{Without Ensemble} & LDA              & $66.7\pm11.7$  & $58.9\pm10.3$              & $\underline{70.4\pm10.4}$ & $68.9\pm9.0$                    & $66.7\pm6.1$  \\ 
                                    & KNN & $\underline{72.2\pm7.9}$   & $54.4\pm8.2$                & $61.4\pm9.3$  & $70.0\pm10.9$                   & $62.2\pm4.2$  \\ 
                                    & AdaBoost         & $\underline{73.3\pm5.4}$  & $48.9\pm8.9$                & $67.9\pm13.5$ & $65.6\pm5.4$                  & $72.2\pm7.9$  \\ 
\multicolumn{2}{c}{With Ensemble}                      & $70.0\pm9.7$     & $53.3\pm10.9$               & $69.0\pm10.8$   & $70.0\pm10.3$                   & $\underline{70.0\pm12.0}$     \\ \hline
\end{tabular}}
\label{tab:machinelearningresultsbinary}
\end{table*}

\section{Discussion and Limitations}

\subsection{User-Centered Design Implications}

This study aims to evaluate the sensitivity of differing psychophysiological dimensions for MWL and PL as well as to build a predictive classification model that differentiates between different levels of MWL and PL. Our framework uses minimally invasive and widely used psychophysiological measurements that exist in modern cars~\cite{2019BMW2019,2020Toyota2020,2021Mercedes-Benz2021} to classify MWL and PL levels towards a user-centered design approach that accommodates the driver's mental capacities. We mimic the driver's interaction with the vehicle's interface while driving through the auditory n-back task and the visual search task, which are equivalent to a speech-based and a touch-based interaction respectively.
While our work focuses on this general dual task approach of manipulating an interface while driving, it is also applicable to more specific applications such as transfer-of-control scenarios for automated and semi-automated vehicles~\cite{odachowska2021psychological}; simplified interfaces, or personalized warnings for mentally exhausting situations~\cite{reyes2020adaptive}; and increasing the user's trust through system awareness and transparency~\cite{hartwich2021improving}. Thus, an adaptive personalized interface observant of the driver's mental capacities can be implemented. For example, high MWL and PL classification can be utilized to simplify the interaction approach with the in-vehicle HMI while driving~\cite{reyes2020adaptive}, or to alert the user to stop driving entirely if needed until mental or perceptual ``cool-down''. 
Finally, while the n-back task is widely used to control MWL levels in the driving context~\cite{Solovey2014classifying,mehler2009impact,Mehler2011AWorkload,Mehler2012SensitivityGroups,ross2014investigating,Barua2020TowardsClassification}, it is limited in regards to construct, concurrent, and ecological validity~\cite{Kane2007WorkingValidity,Jaeggi2010TheMeasure}. Therefore, future studies could investigate other MWL-inducing forms such as traffic or road manipulation ~\cite{Paxion2014MentalDriving,FASTENMEIER2007drivingtaskanalysis}.   

\subsection{Sensitivity of Psychophysiological Measurements }

We found a significant correlation between heart rate and heart rate variability, indicating a convergent validity for heart features. The analysis of the performance measures of the two secondary tasks revealed that the manipulation of three load levels resulted in three difficulty levels each. For the n-back task, there was a significant effect of MWL on the psychophysiological dimensions. This was the case for HR and HRV between the low load and the higher ones. The visual search task showed a non-significant effect for PL for the same dimensions. However, in follow-up tests, there was a significant effect for HR only between the low level and above for PL. 
Similarly, our classification model was able to differentiate between the low MWL (i.e., 1-back task) and the higher ones (i.e., 2 and 3-back tasks), aligning with the statistical analyses. However, it was not able to differentiate between the medium and high MWL levels. 
Next, we further discuss our interpretation and reasoning for these results, as well as the hypotheses.

\subsection{Empirical Hypotheses}
\textbf{H1} expects to find meaningful correlations between the psychophysiological measurements, since they all relate to MWL~\cite{tao2019systematic,Marquart2015ReviewWorkload,Duchowski2020TheActivity}. However, these correlations should be higher for the same measurement type than for different measurement types (i.e., convergent vs. discriminant validity). Since there was only one negative significant correlation between HR and HRV, the convergent validity for heart features is fulfilled~\cite{Matthews2015TheDivergent}. However, there was no significant correlation between the heart features and the eye features, as well as no significant correlation between the eye features among themselves. Besides, the reliability of LHIPA for both eyes was unacceptably low. While ICA shows significant correlations with heart rate dimension in previous studies~\cite{Matthews2015TheDivergent}, LHIPA (a non-patented alternative) fails to do so in our study. This could perhaps be due to light variations as LHIPA is still not thoroughly tested for light invariance and needs further investigation as mentioned by the authors~\cite{Duchowski2020TheActivity}. Moreover, while LHIPA was tested on several controlled environments, it was applied to a much less controlled one in our study (e.g., extreme light variance, head movements, eye movements).

\textbf{H2} states that the psychophysiological dimensions should show differences in MWL.
Due to LHIPA's unreliability, we were able to analyze this hypothesis for heart features only. Both HR and HRV showed significant effects for MWL and were able to distinguish between the low load level compared to higher ones; however, an effect between medium and high load levels could not be identified. While Mehler et al. \cite{Mehler2012SensitivityGroups} were able to identify differences for MWL on three load levels (for heart rate using a similar setup), they used a 0-, 1-, and 2-back task as manipulation in contrast to our study that utilized a 1-, 2-, and 3-back task due to our interest in extremely high load levels. Thus, we hypothesize that in our study, the driver's MWL reached a level of mental saturation such that its effect can no longer be seen in the psychophysiological dimensions for higher load levels~\cite{Wickens2007HowTsang,wickens2008multiple}.

\textbf{H3} states that MWL affects driving performance in the LCT. This hypothesis could not be tested, because the reliability of the driving performance did not meet the minimum acceptable requirements. This could be due to the measured deviation from the reference ideal route not being pronounced enough, or the effect of the mental distraction being just momentary and thus masked in the averaging process. 
As for \textbf{H4}, it relates to the PL effect. For the same previously mentioned reasons, this hypothesis could not be analyzed for the driving performance. As for the psychophysiological measurements, the statistical analysis revealed that there is a non-significant effect for the heart features. 
However, this does not necessarily mean that PL does not influence psychophysiological data. Previous studies showed that other psychophysiological measures (e.g., eye features) can be influenced by the PL of a task~\cite{chen2014using, oliva2019pupil}. 
Since we measured the effects of HR and HRV in a dual task scenario, it could be that having to perform a PL task as a secondary task compared to a single task is already load-inducing for HR and HRV. Therefore, these dimensions are more affected by the difference between driving without a secondary task and driving while using an HMI~\cite{Reyes2008EffectsPerformance}.
Despite that, we found a highly significant effect for mean heart rate when comparing low and higher PL levels (with a high effect size). This might suggest that mean heart rate is affected by PL, especially in lower PL levels. However, since there is a lack of studies analyzing PL effects on heart data, it should also be further investigated in future studies.  

Finally, participants' demographics were limited in our study. Drivers were relatively young ($30.07$ mean age) and experienced ($10.20$ mean driving years). This might have a different effect on their MWL, PL and driving performance compared to older and less experienced drivers~\cite{hakamies1999age,Cantin2009MentalComplexity,Paxion2014MentalDriving}.

\section{Conclusion and Future Work}

Driving remains a complex task that is performed by hundreds of millions of people every day. With the increase in the number of distractors in a car, more personalized and adaptive in-vehicle interfaces are needed. In this paper, we have shown possible approaches to analyze mental and perceptual loads as well as classifying their elevated levels. Elevated load estimation is quite important in anticipating a driver's distraction level, fatigue, stress, and inattentiveness. In our analysis, we were able to show a high correlation between heart rate data and elevation in mental workload. We propose a machine learning based framework for workload level classification through non-intrusive real-time measurements that is technologically ready for use in terms of deployment and applicability. Moreover, our work highlights a potential for further investigation of important factors in driver distraction and cognitive resource management. 

\begin{acks}

This work is partially funded by the German Ministry of Education and Research (BMBF) under project APX-HMI (Grant Number: 01IS17043) and project CAMELOT (Grant Number: 01IW20008).

\end{acks}

\bibliographystyle{ACM-Reference-Format}
\bibliography{references}


\appendix

\section{Correlation Matrices}
\label{sec:appendix_corr}

This appendix shows intra- and inter-correlation matrices for the psychophysiological dimensions.~\autoref{tab:LHIPA} shows the intra-correlations between the left and right eye features (i.e., LHIPA) per secondary task per load level while~\autoref{tab:HRHRV} shows the similar intra-correlations calculations for the heart features (i.e., HR and HRV-RMSSD). Finally,~\autoref{tab:BOTH} shows the inter-correlations across these two psychophysiological dimensions.

\begin{table*}[t]
\caption{Pearson’s correlation between eye features (i.e., LHIPA calculation of the right eye and the left eye) per mental workload level for the n-back task and per perceptual load level for the visual search task. (*) highlights a correlation with $p<.05$ and (**) highlights a correlation with $p<.001$.}
\Description{Person's correlation between eye features per mental workload level for the n-back task and per perceptual load level for the visual search task. The table shows some strong correlations among a few pairs, but they are overall mostly not correlated }
\resizebox{\linewidth}{!}{
\begin{tabular}{ccclllllllllllllll}
\hline
\multicolumn{1}{l}{}           & \multicolumn{1}{l}{}   & \multicolumn{1}{l}{} & \multicolumn{7}{c}{N-Back}                                                                               & \multicolumn{1}{c}{} & \multicolumn{7}{c}{Visual Search}                                                                   \\ \cline{4-10} \cline{12-18} 
\multicolumn{1}{l}{}           & \multicolumn{1}{l}{}   & \multicolumn{1}{l}{} & \multicolumn{3}{c}{LHIPA Right Eye}                     & \multicolumn{1}{c}{} & \multicolumn{3}{c}{LHIPA Left Eye}           & \multicolumn{1}{c}{} & \multicolumn{3}{c}{LHIPA Right Eye}                      & \multicolumn{1}{c}{} & \multicolumn{3}{c}{LHIPA Left Eye}     \\ \cline{4-6} \cline{8-10} \cline{12-14} \cline{16-18}
\multicolumn{1}{l}{}           & \multicolumn{1}{l}{}   & \multicolumn{1}{l}{} & LL               & ML               & HL     &                      & LL     & ML               & HL     &                      & LL                & ML               & HL     &                      & LL              & ML    & HL \\ \hline
\multirow{7}{*}{N-Back}        & \multirow{3}{*}{LHIPA Right Eye} & LL                   & 1                &                  &        &                      &        &                  &        &                      &                   &                  &        &                      &                 &       &    \\
                               &                        & ML                   & 0.268            & 1                &        &                      &        &                  &        &                      &                   &                  &        &                      &                 &       &    \\
                               &                        & HL                   & 0.075            & \textbf{0.601**} & 1      &                      &        &                  &        &                      &                   &                  &        &                      &                 &       &    \\
                               &                        &                      &                  & \textbf{}        &        &                      &        &                  &        &                      &                   &                  &        &                      &                 &       &    \\
                               & \multirow{3}{*}{LHIPA Left Eye} & LL                   & \textbf{0.754**} & 0.080            & -0.118 &                      & 1      &                  &        &                      &                   &                  &        &                      &                 &       &    \\
                               &                        & ML                   & -0.120           & 0.268            & 0.051  &                      & 0.162  & 1                &        &                      &                   &                  &        &                      &                 &       &    \\
                               &                        & HL                   & -0.207           & 0.184            & 0.295  &                      & -0.009 & \textbf{0.471**} & 1      &                      &                   &                  &        &                      &                 &       &    \\
                               &                        &                      &                  &                  &        &                      &        & \textbf{}        &        &                      &                   &                  &        &                      &                 &       &    \\
\multirow{7}{*}{Visual Search} & \multirow{3}{*}{LHIPA Right Eye} & LL                   & -0.027           & -0.101           & 0.222  &                      & 0.053  & 0.022            & 0.126  &                      & 1                 &                  &        &                      &                 &       &    \\
                               &                        & ML                   & -0.054           & -0.024           & 0.182  &                      & 0.098  & 0.125            & -0.025 &                      & -0.021            & 1                &        &                      &                 &       &    \\
                               &                        & HL                   & 0.066            & -0.075           & -0.140 &                      & 0.136  & 0.262            & 0.066  &                      & 0.052             & \textbf{0.436**} & 1      &                      &                 &       &    \\
                               &                        &                      &                  &                  &        &                      &        &                  &        &                      &                   & \textbf{}        &        &                      &                 &       &    \\
                               & \multirow{3}{*}{LHIPA Left Eye} & LL                   & 0.251            & \textbf{0.351*}  & 0.157  &                      & 0.231  & 0.191            & 0.020  &                      & -0.179            & 0.063            & -0.004 &                      & 1               &       &    \\
                               &                        & ML                   & -0.247           & -0.249           & -0.246 &                      & -0.215 & -0.029           & 0.129  &                      & \textbf{-0.466**} & 0.185            & -0.018 &                      & -0.193          & 1     &    \\
                               &                        & HL                   & 0.016            & 0.178            & 0.045  &                      & 0.081  & -0.009           & -0.004 &                      & \textbf{-0.306*}  & -0.031           & -0.009 &                      & \textbf{0.371*} & 0.231 & 1 \\ \hline
\end{tabular}}
\label{tab:LHIPA}
\end{table*}

\begin{table*}[t]
\caption{Pearson’s correlation between heart rate features (i.e., HR and HRV-RMSSD) per mental workload level for the n-back task and per perceptual load level for the visual search task. (*) highlights a correlation with $p<.05$ and (**) highlights a correlation with $p<.001$.}
\Description{Person's correlation between heart rate features per mental workload level fro the n-back task and per perceptual load level for the visual search task. The table shows strong correlations among all pairs}
\resizebox{\linewidth}{!}{
\begin{tabular}{ccclllllllllllllll}
\hline
\multicolumn{1}{l}{}           & \multicolumn{1}{l}{} &    & \multicolumn{7}{c}{N-Back}                                                                                                             & \multicolumn{1}{c}{} & \multicolumn{7}{c}{Visual Search}                                                                                      \\ \cline{4-10} \cline{12-18} 
\multicolumn{1}{l}{}           & \multicolumn{1}{l}{} &    & \multicolumn{3}{c}{HRV-RMSSD}                                & \multicolumn{1}{c}{} & \multicolumn{3}{c}{HR}                                 & \multicolumn{1}{c}{} & \multicolumn{3}{c}{HRV-RMSSD}                                & \multicolumn{1}{c}{} & \multicolumn{3}{c}{HR}                 \\ \cline{4-6} \cline{8-10} \cline{12-14} \cline{16-18} 
\multicolumn{1}{l}{}           & \multicolumn{1}{l}{} &    & LL               & ML               & HL               &                      & LL               & ML               & HL               &                      & LL               & ML               & HL               &                      & LL              & ML              & HL \\ \hline
\multirow{7}{*}{N-Back}        & \multirow{3}{*}{HRV-RMSSD} & LL & 1                &                  &                  &                      &                  &                  &                  &                      &                  &                  &                  &                      &                 &                 &    \\
                               &                      & ML & \textbf{0.875**}  & 1                &                  &                      &                  &                  &                  &                      &                  &                  &                  &                      &                 &                 &    \\
                               &                      & HL & \textbf{0.891**}  & \textbf{0.936**}  & 1                &                      &                  &                  &                  &                      &                  &                  &                  &                      &                 &                 &    \\
                               &                      &    & \textbf{}        & \textbf{}        &                  &                      &                  &                  &                  &                      &                  &                  &                  &                      &                 &                 &    \\
                               & \multirow{3}{*}{HR}  & LL & \textbf{-0.524**} & \textbf{-0.519**} & \textbf{-0.602**} & \textbf{}            & 1                &                  &                  &                      &                  &                  &                  &                      &                 &                 &    \\
                               &                      & ML & \textbf{-0.468**} & \textbf{-0.529**} & \textbf{-0.594**} & \textbf{}            & \textbf{0.942**}  & 1                &                  &                      &                  &                  &                  &                      &                 &                 &    \\
                               &                      & HL & \textbf{-0.445**} & \textbf{-0.496**} & \textbf{-0.602**} & \textbf{}            & \textbf{0.921**}  & \textbf{0.963**}  & 1                &                      &                  &                  &                  &                      &                 &                 &    \\
                               &                      &    & \textbf{}        & \textbf{}        & \textbf{}        & \textbf{}            & \textbf{}        & \textbf{}        &                  &                      &                  &                  &                  &                      &                 &                 &    \\
\multirow{7}{*}{Visual Search} & \multirow{3}{*}{HRV-RMSSD} & LL & \textbf{0.936**}  & \textbf{0.825**}  & \textbf{0.841**}  & \textbf{}            & \textbf{-0.461**} & \textbf{-0.411**} & \textbf{-0.394**} & \textbf{}            & 1                &                  &                  &                      &                 &                 &    \\
                               &                      & ML & \textbf{0.888**}  & \textbf{0.918**}  & \textbf{0.906**}  & \textbf{}            & \textbf{-0.505**} & \textbf{-0.474**} & \textbf{-0.464**} & \textbf{}            & \textbf{0.901**}  & 1                &                  &                      &                 &                 &    \\
                               &                      & HL & \textbf{0.923**}  & \textbf{0.840**}  & \textbf{0.850**}  & \textbf{}            & \textbf{-0.488**} & \textbf{-0.474**} & \textbf{-0.434**} & \textbf{}            & \textbf{0.919**}  & \textbf{0.852**}  & 1                &                      &                 &                 &    \\
                               &                      &    & \textbf{}        & \textbf{}        & \textbf{}        & \textbf{}            & \textbf{}        & \textbf{}        & \textbf{}        & \textbf{}            & \textbf{}        & \textbf{}        &                  &                      &                 &                 &    \\
                               & \multirow{3}{*}{HR}  & LL & \textbf{-0.510**} & \textbf{-0.488**} & \textbf{-0.565**} & \textbf{}            & \textbf{0.941**}  & \textbf{0.907**}  & \textbf{0.884**}  & \textbf{}            & \textbf{-0.511**} & \textbf{-0.511**} & \textbf{-0.511**} & \textbf{}            & 1               &                 &    \\
                               &                      & ML & \textbf{-0.427**} & \textbf{-0.438**} & \textbf{-0.523**} & \textbf{}            & \textbf{0.934**}  & \textbf{0.889**}  & \textbf{0.887**}  & \textbf{}            & \textbf{-0.430**} & \textbf{-0.485**} & \textbf{-0.411**} & \textbf{}            & \textbf{0.960**} & 1               &    \\
                               &                      & HL & \textbf{-0.474**} & \textbf{-0.456**} & \textbf{-0.549**} & \textbf{}            & \textbf{0.909**}  & \textbf{0.904**}  & \textbf{0.889**}  & \textbf{}            & \textbf{-0.462**} & \textbf{-0.492**} & \textbf{-0.527**} & \textbf{}            & \textbf{0.954**} & \textbf{0.930**} & 1  \\ \hline
\end{tabular}}
\label{tab:HRHRV}
\end{table*}

\begin{table*}[t]
\caption{Pearson’s correlation between heart rate features (i.e., HR and HRV-RMSSD) and eye features (i.e., LHIPA calculation of the right eye and the left eye) per mental workload level for the n-back task and per perceptual load level for the visual search task. (*) highlights a correlation with $p<.05$ and (**) highlights a correlation with $p<.001$.}
\Description{Person's correlation between heart rate features and eye features per mental workload level for the n-back task and per perceptual load level for the visual search task. The table shows some correlation between some of the pairs in mostly the middle levels, but overall, not much correlation is seen.}
\resizebox{\linewidth}{!}{
\begin{tabular}{ccclllllllllllllll}
\hline
\multicolumn{1}{l}{}           & \multicolumn{1}{l}{}  &    & \multicolumn{7}{c}{N-Back}                                                                                                    & \multicolumn{1}{c}{} & \multicolumn{7}{c}{Visual Search}                                                                              \\ \cline{4-10} \cline{12-18} 
\multicolumn{1}{l}{}           & \multicolumn{1}{l}{}  &    & \multicolumn{3}{c}{HRV-RMSSD}                          & \multicolumn{1}{c}{} & \multicolumn{3}{c}{HR}                              & \multicolumn{1}{c}{} & \multicolumn{3}{c}{HRV-RMSSD}          & \multicolumn{1}{c}{} & \multicolumn{3}{c}{HR}                               \\ \cline{4-6} \cline{8-10} \cline{12-14} \cline{16-18} 
\multicolumn{1}{l}{}           & \multicolumn{1}{l}{}  &    & LL             & ML             & HL             &                      & LL              & ML              & HL              &                      & LL     & ML             & HL     &                      & LL               & ML              & HL              \\ \hline
\multirow{7}{*}{N-Back}        & \multirow{3}{*}{LHIPA Right Eye} & LL & 0.114          & 0.087          & 0.009          &                      & 0.169           & 0.208           & 0.242           &                      & 0.235  & 0.242          & 0.123  &                      & 0.095            & 0.109           & 0.111           \\
                               &                       & ML & 0.178          & 0.142          & 0.123          &                      & 0.171           & 0.166           & 0.166           &                      & 0.149  & 0.132          & 0.181  &                      & 0.208            & 0.188           & 0.157           \\
                               &                       & HL & 0.201          & 0.187          & 0.157          &                      & 0.086           & 0.103           & 0.063           &                      & 0.172  & 0.158          & 0.160  &                      & 0.148            & 0.100           & 0.161           \\
                               &                       &    &                &                &                &                      &                 &                 &                 &                      &        &                &        &                      &                  &                 &                 \\
                               & \multirow{3}{*}{LHIPA Left Eye} & LL & 0.161          & 0.167          & 0.125          &                      & 0.055           & 0.117           & 0.148           &                      & 0.287  & \textbf{0.336*} & 0.173  &                      & 0.045            & 0.085           & 0.067           \\
                               &                       & ML & 0.056          & 0.067          & 0.098          &                      & \textbf{-0.381*} & \textbf{-0.362*} & \textbf{-0.324*} & \textbf{}            & 0.016  & 0.080          & 0.035  &                      & -0.278           & -0.278          & -0.290          \\
                               &                       & HL & -0.230         & -0.218         & -0.172         &                      & 0.092           & 0.113           & 0.051           &                      & -0.183 & -0.201         & -0.220 &                      & 0.154            & 0.139           & 0.166           \\
                               &                       &    &                &                &                &                      &                 &                 &                 &                      &        &                &        &                      &                  &                 &                 \\
\multirow{7}{*}{Visual Search} & \multirow{3}{*}{LHIPA Right Eye} & LL & -0.134         & -0.160         & -0.128         &                      & 0.035           & 0.035           & 0.017           &                      & -0.152 & -0.184         & -0.160 &                      & 0.146            & 0.068           & 0.112           \\
                               &                       & ML & \textbf{0.322*} & \textbf{0.328*} & \textbf{0.357*} & \textbf{}            & \textbf{-0.381*} & \textbf{-0.306*} & \textbf{-0.337*} & \textbf{}            & 0.230  & \textbf{0.353*} & 0.264  &                      & -0.263           & -0.283          & -0.293          \\
                               &                       & HL & -0.058         & -0.138         & -0.116         &                      & -0.226          & -0.236          & -0.232          &                      & -0.064 & -0.115         & -0.080 &                      & -0.221           & -0.172          & -0.233          \\
                               &                       &    &                &                &                &                      &                 &                 &                 &                      &        &                &        &                      &                  &                 &                 \\
                               & \multirow{3}{*}{LHIPA Left Eye} & LL & 0.084          & 0.053          & -0.012         &                      & 0.060           & 0.150           & 0.201           &                      & 0.159  & 0.101          & 0.127  &                      & 0.026            & 0.015           & 0.055           \\
                               &                       & ML & 0.112          & 0.135          & 0.209          &                      & \textbf{-0.314*} & \textbf{-0.310*} & \textbf{-0.364*} & \textbf{}            & 0.101  & 0.135          & 0.066  &                      & \textbf{-0.399**} & \textbf{-0.368*} & \textbf{-0.355*} \\
                               &                       & HL & 0.215          & 0.247          & 0.221          &                      & -0.050          & -0.089          & -0.108          &                      & 0.253  & 0.282          & 0.187  &                      & -0.104           & -0.068          & -0.063          \\ \hline
\end{tabular}}
\label{tab:BOTH}
\end{table*}

\end{document}